\documentclass[aip,reprint,cha]{revtex4-2}
\usepackage{color}
\usepackage{svg}
\usepackage{amsmath}
\usepackage{amssymb}
\usepackage{amsthm}
\usepackage{bm}
\usepackage{commath}

\theoremstyle{definition}

\theoremstyle{plain}

\theoremstyle{plain}

\theoremstyle{plain}

\theoremstyle{remark}

\theoremstyle{remark}

\usepackage{mathtools}

\begin{document}
	
\title{Synchronization transitions in Kuramoto networks with higher-mode interaction} 

\author{Rico Berner}
\email[]{rico.berner@physik.hu-berlin.de}
\affiliation{Institut f\"ur Physik, Humboldt-Universit\"at zu Berlin, Newtonstraße 15, 12489 Berlin, Germany.}
\author{Annie Lu}
\affiliation{Department of Mathematics, Washington State University, Pullman, Washington, USA.}
\author{Igor M. Sokolov}
\affiliation{Institut f\"ur Physik, Humboldt-Universit\"at zu Berlin, Newtonstraße 15, 12489 Berlin, Germany.}

\date{\today}

\begin{abstract}
Synchronization is an omnipresent collective phenomenon in nature and technology, whose understanding is in particular for real-world systems still elusive. We study the synchronization transition in a phase oscillator system with two nonvanishing Fourier-modes in the interaction function and hence going beyond the Kuromoto paradigm. We show that the transition scenarios crucially depend on the interplay of the two coupling-modes. We describe the multistability induced by the presence of a second coupling-mode. By extending the collective coordinate approach, we describe the emergence of various states observed in the transition from incoherence to coherence. Remarkably, our analysis suggests that in essence the two-mode coupling gives rise to states that are characterized by two independent but interacting groups of oscillators. We believe that these findings will stimulate future research on dynamical systems including complex interaction functions beyond the Kuramoto-type.
\end{abstract}


\maketitle 

\begin{quotation}
Over the last decades, the studies on the collective phenomenon of synchronization in complex dynamical systems have improved our understanding on how different parts of a system can work together seamlessly and efficiently. This is crucial for many everyday systems such as communication networks, power grids and even the human brain. Without synchronization, these systems can experience problems such as data loss, blackouts and neurological disorders. By studying synchronization, we can learn how to design and improve these systems to make them more stable and effective. By studying synchronization, scientists and engineers can develop new techniques to control and improve these systems, leading to better efficiency, stability, and accuracy. In essence, the study of synchronization is a key to unlocking the potential of complex systems and improving our lives. In order to gain a better understanding of real-world dynamical systems of interacting oscillators, studying realistic interaction functions has to be accessible by existing mathematical methods. In this work, we provide a next step into this direction and provide detailed insights into the synchronization transition for systems of coupled phase oscillators with a coupling beyond the Kuramoto-type.
\end{quotation}
\section{Introduction}\label{sec:intro}
Dynamical networks of phase oscillators are a commonly used paradigm for studying the synchronization patterns in systems of interacting agents~\cite{PIK01,ACE05}. System of weakly interacting nonlinear oscillators can be generally reduced to a network of phase oscillators~\cite{WIN80,HOP97,PIK01,PIE19a} which makes this class of models an even more important one when it comes to the description of collective phenomena. The importance of phase oscillator models and reduction techniques have been highlighted in various reviews and books~\cite{PIK01,ASH16,BIC20}. Recent studies also aim at increasing the range of applicability of phase oscillator models by generalizing the conditions under which reduction techniques are valid~\cite{KLI17a,ERM19,ROS19a,ROS19b,PIK22}.

A famous representative of the class of phase oscillator models is the Kuramoto model where all oscillators are coupled in the "all-to-all" manner. Due to its simple form and mathematical tractability this model has attracted much attention~\cite{KUR84,STR00}. Over the last decades several extensions of the Kuramoto model have gained additional popularity through applications to real-world problems~\cite{STR93,PIK01,STR03,ROD16}, including neuroscience~\cite{BRE10h,LUE16,ASL18a,ROE19a,BIC20}, physiological models~\cite{SAW21c,BER22} or power grids~\cite{FIL08a,HEL20,TOT20,BER21a}. Moreover, the Kuramoto model has been extended to study synchronization on static~\cite{BOC16,ROD16,BOC18}, temporal~\cite{GHO22} and adaptive networks~\cite{SEL02,MAI07,AOK12,KAS17,BER19,BER20,BER21c,VOC21}. Despite the simple structure, extended Kuramoto models can exhibit many different dynamical regimes such as solitary~\cite{MAI14a,TEI19,TAH19,BER20c} and chimera states~\cite{OME13,OME19c,PAR21a,BER21b,THA22}, and sophisticated methods have been developed for their analysis~\cite{OME18a}. 

The collective dynamics of coupled phase oscillator systems that meet certain requirements on the frequency distribution and the coupling structure have been extensively analyzed. For this dimensional reduction techniques as the Watanabe-Strogatz theory~\cite{WAT93a,WAT94,MAR09c,STE11b} and the Ott-Antonson ansatz~\cite{OTT08} have been utilized. More recently another method with relation to the Ott-Antonsen approach has been developed based on a Galerkin approximation~\cite{GOT15,SMI20}. This so-called collective coordinate approach describes the synchronization or chaotic cluster dynamics in generalized Kuramoto systems. Different to the other methods, the collective coordinate approach can be applied with less restrictions, however, the accuracy of the approach depends on the proper choice of an ansatz manifold. The collective coordinate approach has been recently successfully applied to phase oscillator systems with multimodal frequency distributions~\cite{SMI19}, with complex coupling structure~\cite{HAN18b,YUE20,SMI21}, or with adaptive coupling weights~\cite{FIA22}.

Phase transitions play an important role in natural sciences~\cite{STA71} and are a commonly studied phenomenon in systems of coupled heterogeneous oscillators. Complex dynamical networks~\cite{NEW03,BOC18} exhibit a plethora of nonequilibrium phase transitions describing changes in collective dynamics in response to variations in control parameters such as interaction strength. In particular, transitions between coherence and incoherence have attracted significant attention~\cite{ROD16}. Also here, the Kuramoto model~\cite{KUR84} has served as a test bed to study phase transitions in networks of coupled oscillators. It is known that the Kuramoto model exhibits either first or second-order phase transitions from incoherence to full synchronization, depending on the natural frequency distribution~\cite{KUR84,PIK01,PAZ05a,GOM11a,BOC16,SOU19}. Also the network structure~\cite{GOM07} or the weight distributions~\cite{ZHA13a} can have strong effects on the nature of the synchronization transitions.

Even though Kuramoto-like models and the related findings are fundamental for the research on complex dynamical networks, the majority of works is limited to interaction functions consisting of a single mode. While it has been shown that interaction functions with pure higher-mode coupling can be treated similarly as their first-mode counterparts~\cite{DEL19,GON19a}, the dynamics in the presence of a mixed-mode interaction function remains elusive. It has been known that mixed-mode couplings may play an important role in the dynamics of neuronal systems~\cite{HAN93b,TAS99,BRE10h}. The presence of multiple modes can have a significant influence on the dynamics of such systems~\cite{DAI95, DAI96,HAN93a,ASH07,EYD17}. In Refs.~\onlinecite{KOM13a,KOM14}, Komarov and collaborators have found that an additional second mode in the interaction function leads to the emergence of novel steady states in heterogeneous systems. A recent study on Kuramoto-Sakaguchi models of rotators, also called models with inertia, has shown that additional Fourier-modes in the interaction function may lead to tripod patterns named cyclops states that govern the full system's dynamics~\cite{MUN23}. The progress on the phase reduction beyond the weak coupling limit suggests that mixed-mode couplings appear quite naturally in phase model approximations~\cite{ASH16b,LEO19,GEN20}. While synchronization transitions are well studied for single-mode interaction functions, mixed-mode interactions have been investigated only rarely~\cite{LI14b,WAN17e}.

In this work, we study the synchronization transition in a heterogeneous system of all-to-all coupled phase oscillators with a mixed-mode interaction function. In particular, we consider a interaction function consisting of a first and second Fourier mode. The heterogeneity is introduced by a uniform distribution of the oscillators natural frequencies. Depending on the coupling parameters, corresponding to the first or second mode, the systems undergo different transitions to synchrony. We describe two different scenarios of first-order transitions to full synchrony where the first scenario features a cascade of first-order transitions through coexisting two-cluster phase-locked states while the other scenario shows a first-order transition from an anti-phase phase-locked to an in-phase synchronous state. In this work, we show that the zoo of possible transition scenarios is increased by the competing interplay of the first and second mode of the interaction function leading to a plethora of two-cluster phase locked states. Extending the scope of the collective coordinate approach, we derive analytic conditions for the existence of the clustered phase-locked states observed in the transition scenarios.

In the next section, Sec.~\ref{sec:model}, we introduce the model considered throughout the study. Subsequently in Sec.~\ref{sec:transitions}, two very different transition scenarios depending on the form of the interaction function are described an the numerically observed states are discussed. For these states, we develop in Sec.~\ref{sec:collectiveCoord} existence conditions based on the collective coordinate approach. All results are summarized in the concluding Sec.~\ref{sec:conclusion}.

\section{Coupled phase oscillator models and mixed-mode coupling}\label{sec:model}

We consider a heterogeneous system of $N$ coupled phase oscillators that can be generally written as
\begin{align}\label{eq:phaseoscModel_general}
    \frac{\mathrm{d}}{\mathrm{d}t}\bm{\phi} & = \bm{\omega} + F(\bm{\phi}),
\end{align}
where $\bm{\phi}=(\phi_1,\dots,\phi_N)^T$, and each oscillator is represented by a dynamical variable $\phi_j(t)\in[0,2\pi)$, $j=1,\dots,N$. The oscillators possess their individual natural frequency $\omega_i$ that are collected in the vector $\bm{\omega}=(\omega_1,\dots,\omega_N)^T$. Moreover,  $F=(f_1(\bm{\phi}),\dots,f_N(\bm{\phi}))^T$ is the interaction vector field with interaction functions $f_j$ which are assumed to be $2\pi-$periodic.

To measure the phase coherence, we define the $m$th moment of the complex mean field, $m \in \mathbb N$, as
\begin{align}\label{eq:defOrderParam}
	Z_m(\bm{\phi}) = \frac{1}{N}\sum_{j=1}^{N}e^{\mathrm{i}m\phi_j} = R_m(\bm{\phi})e^{\mathrm{i}\rho_m(\bm{\phi})},
\end{align}
where $\mathrm{i}$ is the imaginary unit, $R_m$ denotes the $m$th moment of the (Kuramoto-Daido) order parameter, and $\rho_m$ is the collective phase of the $m$th moment of the mean field~\cite{KUR84,DAI94}.

Throughout this article, we assume that the interaction functions are of the following form where we consider only the first two modes of a Fourier expansion
\begin{align*}
	f_j(\bm{\phi}) = \frac{K_1}{2} Z_1 e^{-\mathrm{i}\phi_j} + \frac{K_2}{2} Z_2 e^{-\mathrm{i}2\phi_j} + \mathrm{c.c.},
\end{align*}
such that we finally arrive at the two-mode model
\begin{align}\label{eq:twoModeModel}
    \frac{\mathrm{d}}{\mathrm{d}t}\phi_j = \omega_j + K_1 \mathrm{Im}(Z_1 e^{-\mathrm{i}\phi_j}) + K_2 \mathrm{Im}(Z_2 e^{-\mathrm{i}2\phi_j})
\end{align}
with coupling strengths to the first and second moded $K_1$ and $K_2$, respectively.
Further, we assume that the natural frequencies $\omega_j$ are drawn from a uniform distribution with distribution density $d(\omega)=1/2a$ for $\omega\in[-a,a]$ with $a\in\mathbb{R}$ and $d(\omega)=0$ otherwise.

A solution of the phase oscillator system~\eqref{eq:phaseoscModel_general} is called  \textit{phase-locked state} if
\begin{align}\label{eq:phaseLockedStates}
	\phi_j(t) = \Omega t + \vartheta_j,\quad i=1,\dots,N,
\end{align}
with collective frequency  $\Omega\in\mathbb{R}$ and fixed relative phases $\vartheta_j\in [0,2\pi)$ of the individual oscillators.
A phase-locked state can be characterized by the corresponding $m$th moments of the order parameter. We call a state an $m$-synchronous state if $R_m=1$ and an $m$-splay state~\cite{BER21e}, also called incoherent state~\cite{STR91}, if $R_m=0$. States with $R_1=1$ and $R_2=1$ are also called in-phase and antipodal states in the literature~\cite{BER19}. These states are respectively described by $\vartheta_j=0$ for all $j=1,\dots,N$ (in-phase) and by two phase-clusters with $\vartheta_j=0$ for $j=1,\dots,N_1$ and $\vartheta_j=\pi$ for $j=N_1+1,\dots,N$ with $1\le N_1<N$ (antipodal). Note the relation between the different moments of the order parameter $R_m(\bm{\phi})=R_1(m\bm{\phi})$ and the fact that the class of antipodal states includes in-phase states. We further note that the in-phase state and the antipodal state with $N_1=N/2$ solve system~\eqref{eq:twoModeModel} in case of equal frequencies, i.e., $\omega_j=0$ and for any $K_1$ and $K_2$. For heterogeneous frequencies, however, phase-locked states, if they exist, will appear perturbed and hence the order parameters will be only approximately $R_{1,2}\approx 1$. 

Despite the fact that distributed natural frequencies disturb the appearance of in-phase and antipodal state, they describe important classes of solution families in the $(K_1,K_2)$-parameter spaces. The latter fact can be seen as follows. Consider $K_2$ ($K_1$) as fixed and the limit $K_1\to\infty$ ($K_2\to\infty$). Then equation determining the steady states of~\eqref{eq:twoModeModel} can be written as $0=\mathrm{Im}(Z_1 e^{-\mathrm{i}\phi_j})$ ($0=\mathrm{Im}(Z_2 e^{-\mathrm{i}2\phi_j})$) where we used $\omega_j/K_1\to 0$ and $K_2/K_1\to 0$ ($\omega_j/K_2\to 0$ and $K_1/K_2\to 0$) and hence the in-phase (antipodal) states are solution in this limit. Note that these two types of states give rise to two families of states for finite values of the coupling strengths $K_1$ or $K_2$. Therefore, we call phase-locked states of in-phase or antipodal-type if they belong the corresponding families, i.e., $\phi_j=0+\mathcal{O}(1/K)$ ($K=K_1$ or $K=K_2$) for all $i=1,\dots,N$, or $\phi_j=0+\mathcal{O}(1/K_2)$ for $j=1,\dots,N_1$ and $\phi_j=\pi+\mathcal{O}(1/K_2)$ for $j=N_1,\dots,N$, respectively.

The limiting cases can be also seen via writing the state state equation for~\eqref{eq:twoModeModel} as 
\begin{align*}
    \frac{\omega_j}{R} = -\left(\frac{R_1 K_1}{R}+\frac{R_2 K_2}{R}\cos \phi_j \right)\sin\phi_j
\end{align*}
with $R=\sqrt{K_1^2 R_1^2 + K_2^2 R_2^2}$ where we assumed that $d(\omega)$ is symmetric in order to set $\rho_1=\rho_2=0$, compare to~[\onlinecite{KOM13a}]. If we consider the limiting cases $K_1\to\infty$ and $K_2\to\infty$, the latter equation reduces to $0=\sin{\phi_j}$ and $0=\cos(\phi_j)\sin{\phi_j}$, respectively.

In the following section, we show the results of a numerical analysis. In order to characterize the dynamical states, we make use of temporally averaged order parameters that are defined as:
\begin{align}\label{eq:avOrderParam}
	\langle R_m\rangle = \frac1T \int_{t_0}^{t_0+T} \left|Z_m(\bm{\phi}(t))\right|\,\mathrm{d}t.
\end{align}
Here, the averaging time window $T$ is considered to sufficiently large and $t_0$ is chosen sufficiently large to neglect transient dynamics.
\section{Transition scenarios between incoherence and coherence in a two-mode model}\label{sec:transitions}
In the previous section, we have introduced the $2$-mode model and have already provided some approximations for phase-locked states in the large coupling limits, i.e., either $K_1\to\infty$ or $K_2\to \infty$. In this section, we show how the system dynamics changes depending on the coupling strengths $K_1$ and $K_2$. In particular, we analyze transitions between states with respect to $K_1$ for two different values of $K_2$. The uniform natural frequency distribution is considered to have its support on $[-1,1]$ throughout the manuscript.

\begin{figure}[h!]
    \centering
    \includegraphics{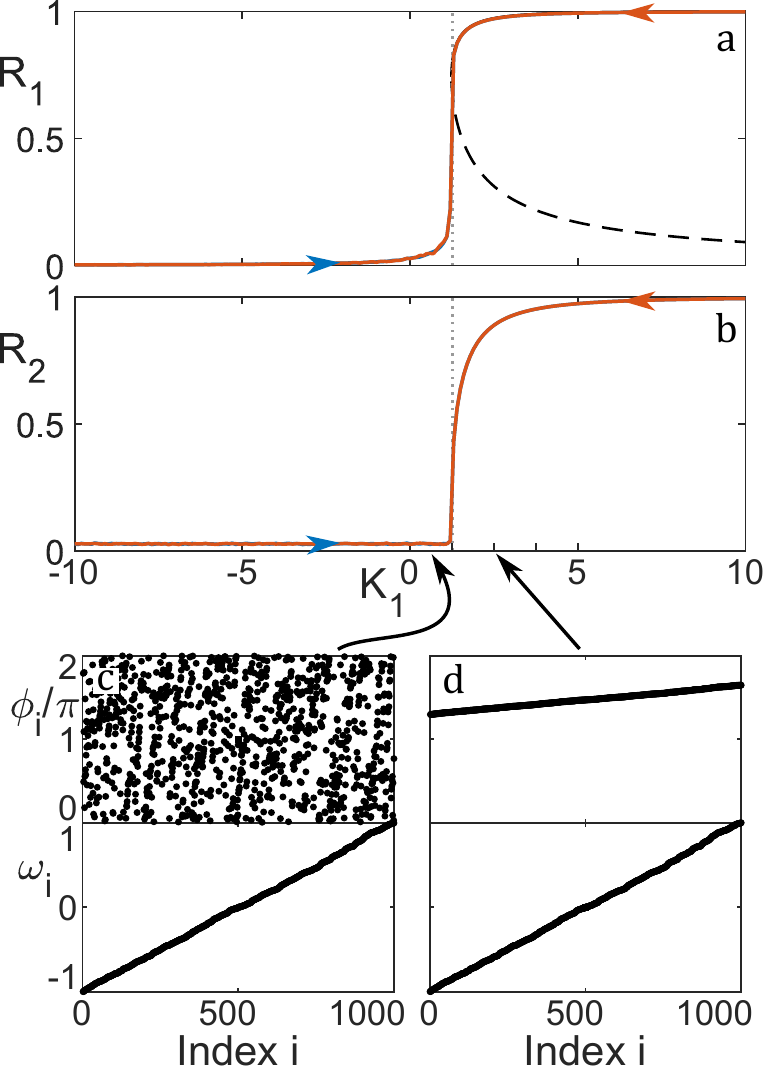}
    \caption{Adiabatic continuation in the coupling strength of the first mode $K_1$. In panels (a,b) we show the results of an adiabatic continuation in $K_1$ for the fixed value $K_2=0$ where we plot the temporally averaged (see Eq.~\ref{eq:avOrderParam}) first moment of the order parameter $R_1$ in (a) and the second moment $R_2$ in (b). For the averaging, we discard the transient time. In the figures, we show the results of the continuous sweep-up (blue) and sweep-down (red) with step size $\Delta K_1=0.1$ starting and ending as indicated in the $K_1$-axis. The direction of the sweeps are indicated by arrows. The gray dotted line indicate the threshold $4/\pi$ beyond which the incoherent state becomes linearly unstable in the continuum limit. The black curve in (a) shows the analytic line of existence  for synchronous states from the collective coordinate approach derived in Sec.~\ref{sec:collectiveCoord}, see Eq.~\eqref{eq:oneClSol}, dashed and solid parts indicate unstable and stable states, respectively. Snapshots of states from the adiabatic continuation are presented in panels (c) and (d). They show the phases and their corresponding natural frequencies sorted with respect to an increasing order of the natural frequencies. Panels (a) and (b) show results for $K_1=0.5$ and $K_1=2$, respectively. Simulation details: In all panels $\omega_i$ are taken from the same uniform distribution on $[-1,1]$, the simulation time is $400$ time units of which $200$ time units. are considered as transient time, for the continuation an increment of $\Delta K_1=0.1$ is considered, $K_2=0$ and $N=1000$.}
    \label{fig:adiabatic_k20}
\end{figure}

This work is concerned with the interaction of two modes in the interaction function, we show for comparison the transition behavior of a system with only single mode interaction function, i.e., $K_2=0$. The results of an adiabatic continuation in $K_1$ are presented in Fig.~\ref{fig:adiabatic_k20}. For the sweep-up (sweep-down) continuation, we simulate system~\eqref{eq:twoModeModel} for a fixed parameter value of $K_1$ and a given time interval, then increase (decrease) the value with increment $\Delta K_1$ and start a new simulation where the initial conditions are given by the final state of the preceding simulation. The sweep-up is started from a random initial condition while the sweep-down is started from the final state of the final simulation of the sweep-up.

\begin{figure}[h!]
    \centering
    \includegraphics{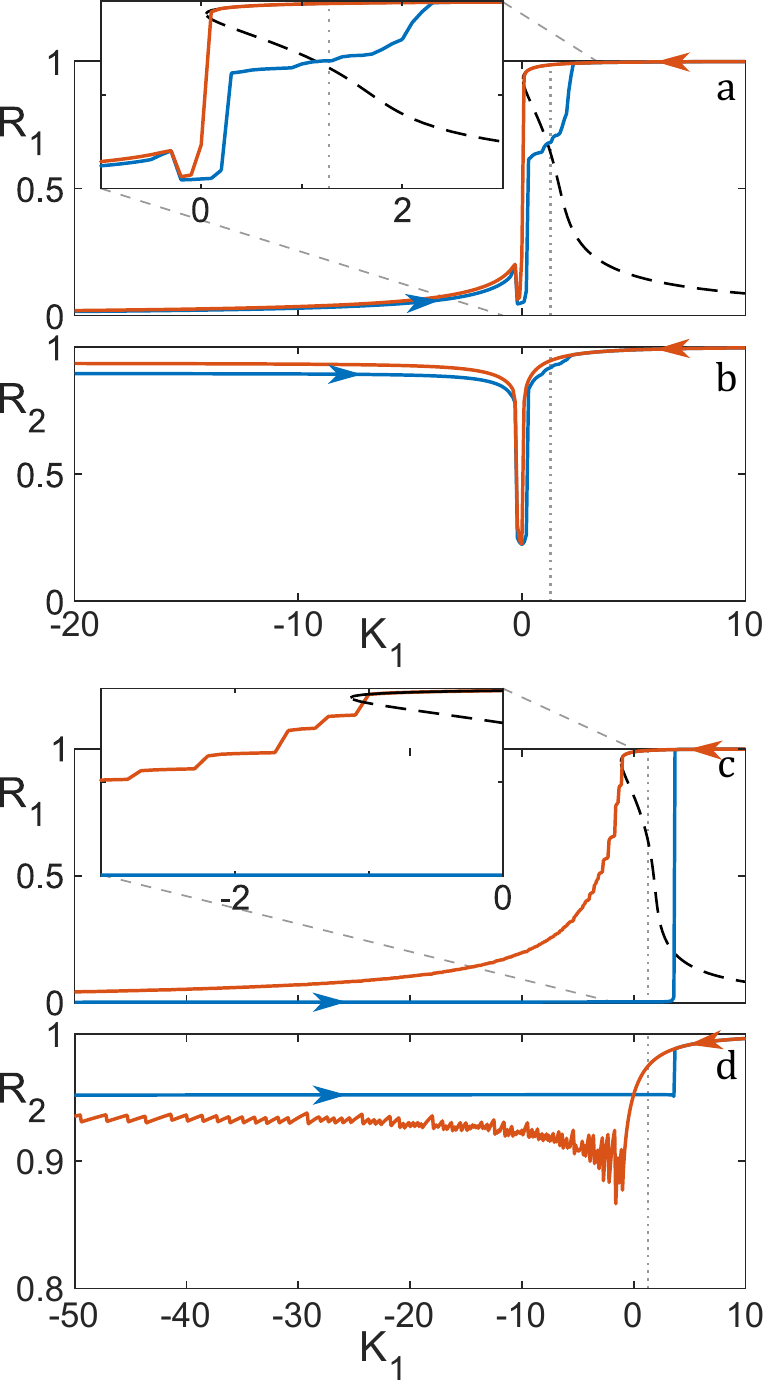}
    \caption{Adiabatic continuation in the coupling strength of the first mode $K_1$. In panels (a,b) and (c,d) we show the results of an adiabatic continuation in $K_1$ for the fixed values $K_2=1.2$ and $K_2=2$, respectively where we plot the temporally averaged (see Eq.~\ref{eq:avOrderParam}) first moment of the order parameter $R_1$ in (a,c) and the second moment $R_2$ in (b,d). For the averaging, we discard the transient time. In the figures, we show the results of the continuous sweep-up (blue) and sweep-down (red) with step size $\Delta K_1=0.1$ starting and ending as indicated in the $K_1$-axis. The directions of the sweeps are indicated by arrows. We provide blow-ups of the transition regions in panels (a) and (c). The gray dotted line indicate the threshold $4/\pi$ beyond which the incoherent state becomes linearly unstable in the continuum limit. The black curve in (a) and (c) shows the analytic line of existence for synchronous states from the collective coordinate approach derived in Sec.~\ref{sec:collectiveCoord}, see Eq.~\eqref{eq:oneClSol}, dashed and solid parts indicate unstable and stable states, respectively. All other parameters as in Fig.~\ref{fig:adiabatic_k20}}
    \label{fig:adiabatic}
\end{figure}

In accordance with the literature~\cite{PAZ05}, we observe a sudden (first-order) transition from complete desynchronization, i.e., incoherent state (Fig.~\ref{fig:adiabatic}c) to full synchronization, i.e, coherent state (Fig.~\ref{fig:adiabatic}d) at the critical value $K_{1,c}=2/\pi d(0)$ where $d(0)$ is the value of the distribution density $d$ at zero. The critical value of $K_1$ determines the point above which the incoherent state becomes linearly unstable in the continuum limit ($N\to\infty$)~\cite{STR91}. In case of a uniform frequency distribution, this critical value also coincides with the saddle-node bifurcation point for the synchronous state\cite{PAZ05}.

\begin{figure}[h!]
    \centering
    \includegraphics{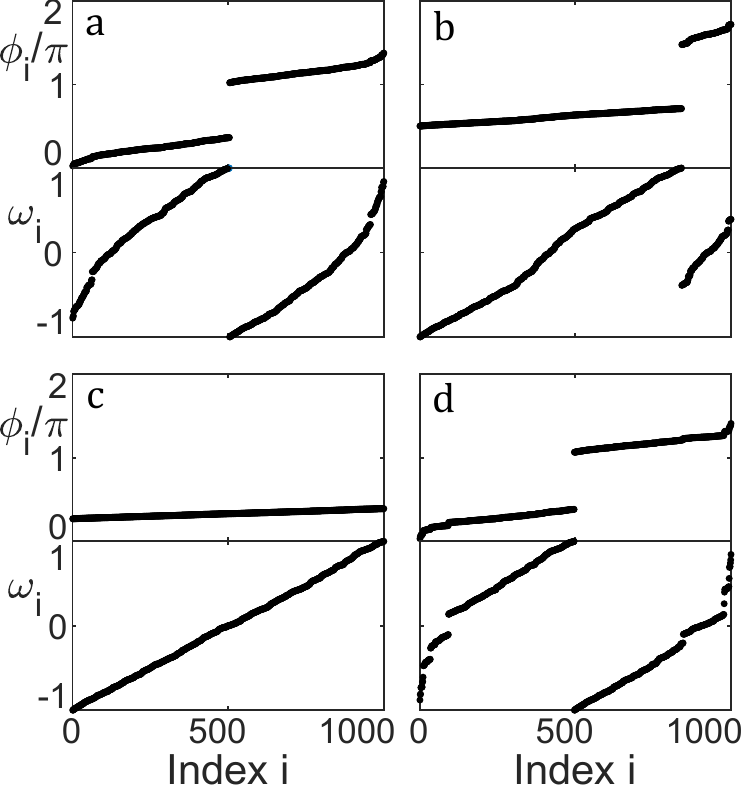}
    \caption{Snapshots of states from the adiabatic continuation shown in Fig.~\ref{fig:adiabatic}(a,b) with $K_2=1.2$ for the sweep-up and sweep-down at different values of $K_1$. The panels show the phases and their corresponding natural frequencies sorted with respect to an increasing order of the phases. Panels (a) and (b) show results from the sweep-up at $K_1=-1$ and $K_1=1.5$ and panels (c) and (d) show results from the sweep-down at $K_1=3$ and $K_1=-1$, respectively.}
    \label{fig:snaps_k2_12}
\end{figure}

\begin{figure}[h!]
    \centering
    \includegraphics{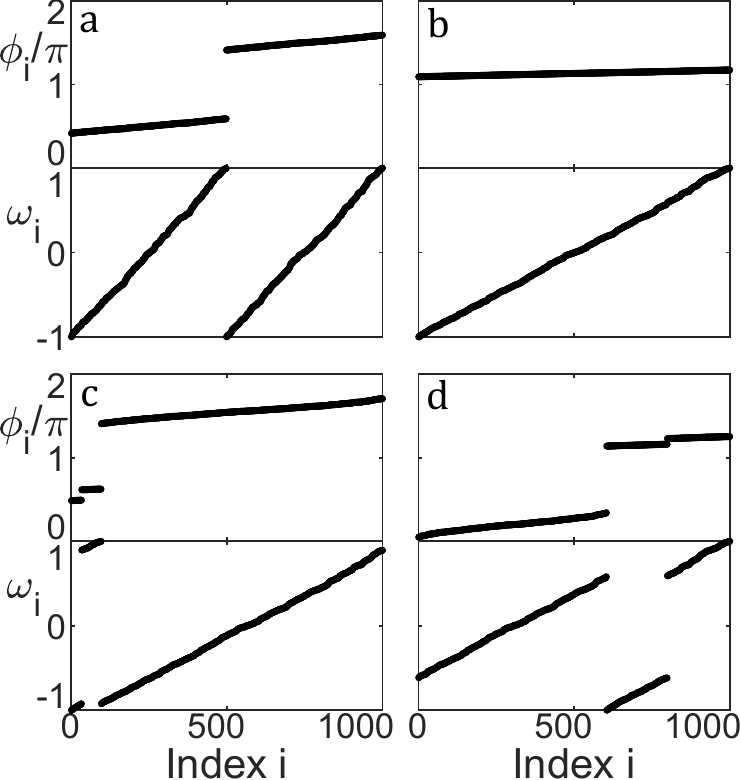}
    \caption{Snapshots of states from the adiabatic continuation shown in Fig.~\ref{fig:adiabatic}(c,d) with $K_2=2$ for the sweep-up and sweep-down at different values of $K_1$. The panels show the phases and their corresponding natural frequencies sorted with respect to an increasing order of the phases. Panels (a) and (b) show results from the sweep-up at $K_1=-10$ and $K_1=4$ and panels (c) and (d) show results from the sweep-down at $K_1=-1.5$ and $K_1=-10$, respectively.}
    \label{fig:snaps_k2_2}
\end{figure}

In Figure~\ref{fig:adiabatic}, we show the results of an adiabatic continuation in $K_1$ for nonvanishing values of $K_2$. Some snapshots of simulations are presented in Fig.~\ref{fig:snaps_k2_12} and \ref{fig:snaps_k2_2} for $K_2=1.2$ and $K_2=2$, respectively. For each state we show the phase distributing at the end of the simulation, sorted in increasing order, along with the corresponding natural frequencies sorted accordingly.

The sweep-up continuation for $K_2=1.2$ presented in Fig.~\ref{fig:adiabatic}(a,b) (blue line, increasing $K_1$) shows the emergence of a phase-locked state at the beginning of the sweep-up at $K_1=-20$. This state is characterized by an almost vanishing average first order parameter $R_1$ and an average second order parameter $R_2$ close to $1$. Fig.~\ref{fig:snaps_k2_12}(a), we provide a snapshot for this state showing that the set of oscillators is split up into two groups that are related by a phase shift $\pi$. We find further that the number of oscillators in the two-cluster of antipodal-type are almost equal ($n_1\approx 0.5$). Looking at the corresponding natural frequencies of the oscillators in each group, we observe a splitting of the uniform distribution on $[-1,1]$ into two uniform distributions on the same interval. With increasing coupling strength $K_1$ this two-cluster state persists until close to $K_1=0$ where the two-cluster disappears and partial cluster states emerge. The partial cluster states bifurcate with increasing coupling strength into a cascade of two-cluster states. One of these two-cluster states emerging in the cascade is shown in Fig.~\ref{fig:snaps_k2_12}(b). We note that the second cluster is much smaller than the first cluster and also the corresponding natural frequencies are uniformly distributed on an interval with tighter bounds. During the cascade the size of the second cluster as well as the size of the distributions interval decreases sequentially. Ultimately, the cascade ends in an one-cluster phase-locked state with $\langle R_{1,2}\rangle\approx 1$ that is persistent for higher values of $K_1$, see Fig.\ref{fig:snaps_k2_12}(c).

The down-sweep continuation starting with the last state of the up-sweep at $K_1=10$ is presented in Fig.~\ref{fig:adiabatic}(a,b) (red line, decreasing $K_1$). With decreasing $K_1$, we see that the one-cluster state can be observed even in the coupling strength range where two-cluster states are found during the sweep-up giving rise to the existence of a multistable regime and the presence of hysteresis. The blow-up in Fig.~\ref{fig:adiabatic}(a) shows that the one-cluster state exists until $K_1\approx 0$ and changes into state before forming a two-cluster state very similar to the one found during the sweep-up, see Fig-\ref{fig:snaps_k2_12}(d). Note that distribution of the natural frequencies is split up slightly differently for the two-cluster states observed in the sweep-up than in the sweep-down.

In Fig.~\ref{fig:adiabatic}(c,d), we analyze the transitions between the collective states for another value of the coupling strength $K_2=2$. Here, in the beginning of the sweep-up, we observe a similar two-cluster state as in the previous case for $K_2=1.2$, see Fig.~\ref{fig:snaps_k2_2}(a), however, with $\langle R_1\rangle$ almost exactly at $0$. This state is very persistent and its form does not change with increasing $K_1$. Only at larger values of $K_1$, this two-cluster state of antipodal-type loses its stability which leads to an abrupt transition to one-cluster phase-locked state, see Fig.~\ref{fig:snaps_k2_2}(b) that stays for higher values of $K_1$. As in the example for $K_2=1.2$, we also observe hysteresis in the sweep-down. Moreover, the one-cluster state exists for values of $K_1$ even below $0$. The disappearance of the phase-locked one-cluster state leads to a cascade of three-cluster states for decreasing $K_1$. In Fig.\ref{fig:snaps_k2_2}(c,d), we show two snapshots representing states from this cascade. Close to the bifurcation point of the one-cluster state, the largest cluster of the emerged three-cluster state is rather big compared to the others, see Fig.\ref{fig:snaps_k2_2}(c). With decreasing $K_1$ the largest cluster shrinks and its relative size tends to $n_1=0.5$, see see Fig.\ref{fig:snaps_k2_2}(d). Further noteworthy, the distribution of natural frequencies is split up into approximately three uniform distributions each on a separate interval $I_1,I_2,I_3$ such that $\bigcup_{i\in{1,2,3}} I_i = [-1,1]$ and where the interval in the middle ranging symmetrically from a negative to positive value belongs to the largest cluster. The cascade of three-cluster states leads to a rather smooth transition from a state with $R_1\approx 1$ for $K_1>0$ to a state with $R_1\approx 0$ for $K_1\to -\infty$.

%
%
\section{Collective coordinate approach for clustered states\label{sec:collectiveCoord}}
As described in section~\ref{sec:model}, in case of multi-mode coupling only little is known on the transition to coherence, in particular, if the coupling strengths $K_1$ and $K_2$ are of the same order. This lack of knowledge is basically due to the fact that well established methods such as the Ott-Antonsen theory are not applicable. In order to overcome some of these issues and to provide analytic insights into the phase-locked states found in the previous section we make use of the collective coordinate approach~\cite{GOT15}. The collective coordinate approach has shown to be related to the Ott-Antonsen approach for an infinite number of phase oscillators~\cite{YUE20,SMI20a}. Beyond this, the method provides very good approximation for coherent states where other methods fail, e.g. for systems with a complex network structure~\cite{HAN18b}, on sparse networks~\cite{SMI21} or with adaptive coupling weights~\cite{FIA22}.

\subsection{General collective coordinate approach}
In the following sections, we use the collective coordinate approach in order to describe the emergence of phase-locked state while it has been used for partially locked states as well~\cite{SMI19,SMI20a}. In this section, we introduce the main methodology that is then adapted for the purposes shown in the subsequent sections. 

For doing so, we consider a general ansatz function $\hat{\bm{\phi}}$ with which we approximate the dynamics of each oscillator as $\phi_i(t)=\hat{\phi}_i(\bm{\alpha}(t);\omega_i)$. The function $\hat{\bm{\phi}}_i$ is called shape function where the shape depends on the $M$ functions $\bm{\alpha}(t)=(\alpha_1(t),\dots,\alpha_M(t))$ called the collective coordinates, and the natural frequencies of the phase oscillators $\omega_i$. Here one might consider a one-dimensional vector $\alpha(t)$ at any time as done in Ref.~\onlinecite{GOT15}. In case of a phase-locked state this assumption is reasonable since the state of all phases is determined by knowing the state of a single reference phase. However, as discussed in the subsequent sections, for mixed-mode interaction the single collective coordinate assumption might be not sufficient. By using the collective coordinate approach, we consequently assume that the solution of the dynamical system of coupled phase oscillators~\eqref{eq:phaseoscModel_general} stays close to an $M$ dimensional manifold whose shape depends on the frequency distribution. In order to obtain a quantitatively good approximation the choice of the ansatz function is crucial.

In case of a single mode coupling, i.e., $K_1=0$ or $K_2=0$, a linear ansatz of the form $\hat{\phi}_i=\alpha(t)\omega_i$ can be motivated by linearizing the general solution for $\phi_i^*$ of $0=\omega_i + K_1 R_1\sin(\rho_1-\phi_i^*)$ if $K_2=0$ or of $0=\omega_i + K_2 R_2\sin(\rho_2-2\phi_i^*)$ if $K_1=0$, see~\cite{SMI20a}. In case of nonvanishing coupling strengths, the shape function might be chosen differently e.g. to account for multiple clusters. We discuss such approaches in the following sections.

Once a shape function has been fixed, the dynamics on the submanifold described by this function has to be determined. For this, we consider the error $\varepsilon_i$ made by reducing the full system to the submanifold
\begin{align*}
    \varepsilon_i = \sum_{m=1}^M \dot{\alpha}_m\frac{\partial\hat{\phi}_i}{\partial\alpha_m} - \dot{\phi}_i(\hat{\phi}_i(\bm{\alpha}(t);\omega_i))
\end{align*}
where $\dot{\phi}_i$ is determined by model~\eqref{eq:twoModeModel}. To minimize the error we impose the condition that $\bm{\varepsilon}=(\varepsilon_1,\dots,\varepsilon_N)^T$ is orthogonal to the tangent space of the submanifold spanned by the vectors $\mathrm{d}\hat{\bm{\phi}}/\mathrm{d}\alpha_m$. Using the Euclidean scalar product, the condition
\begin{align}\label{eq:collCooErrMin}
    \left \langle \bm{\varepsilon},\frac{\mathrm{d}\hat{\bm{\phi}}}{\mathrm{d}\alpha_m} \right\rangle = 0
\end{align}
for $m=1,\dots,M$ yield equations of motion for the collective coordinates $\alpha_m(t)$ of the form
\begin{align}\label{eq:dyn_coll_coord}
    \dot{\alpha}_m = g_m(\bm{\alpha}(t); \bm{\omega})
\end{align}
where $g_m$ results from solving $\eqref{eq:collCooErrMin}$. In the following, we derive the reduced equation for clustered phase-locked states in different collective coordinates. We note at this point that the collective coordinate approach uses prior knowledge of the system's dynamics in order to provide a reasonable approximation. Further, for a given collective coordinate ansatz there might be solutions to the reduced equations that do not properly or at all approximate the actual system's behavior.

\subsection{Approximation of phase-locked states of in-phase-type}\label{sec:collCoordInPhase}
For the the type of phase-locked states that could be regarded as a perturbation from the in-phase synchronous state, i.e., phase-locked state of the in-phase-type, we make use of a linear ansatz for the collective coordinate approach. Therefore, let the shape function by given as~\cite{SMI20a}
\begin{gather}
    \hat{\phi}_i(t) = \alpha(t)\omega_i
\end{gather}
for which the error results in
\begin{align}\label{eq:oneCluster_error}
    \varepsilon_i = \dot{\alpha} \frac{d\hat{\phi}_i}{d\alpha} - \omega_i - K_{1} \mathrm{Im}(Z_1 e^{-\mathrm{i}\hat{\phi}_i}) - K_{2} \mathrm{Im}(Z_2 e^{-\mathrm{i}2\hat{\phi}_i})
\end{align}
with
\begin{gather*}
    Z_m = \frac{1}{N}\sum_{j= 1}^N e^{\mathrm{i}m \alpha(t)\omega_i}
\end{gather*}
for $m=1,2$. Using this ansatz we can reduce the $N$-dimensional two-mode model~\eqref{eq:twoModeModel} to an one-dimensional ordinary differential equation for $\alpha(t)$ by minimizing the error $\bm{\varepsilon}$ as outlined in the section above. Using the condition~\eqref{eq:collCooErrMin} and noting that $\frac{d\hat{\phi}_i}{d\alpha} = \omega_i$, we can arrive at the desired evolution equation for a finite size system in terms of the collective coordinate $\alpha$,
\begin{multline}\label{eq:oneClSol_finite}
    \dot{\alpha} = 1 + \frac{K_{1}}{\Sigma_\omega} \frac{1}{N^2}\sum_{i = 1}^N \omega_i \sum_{j = 1}^N  \sin{\alpha[\omega_j - \omega_i]}  \\
    +\frac{K_{2}}{\Sigma_\omega} \frac{1}{N^2}\sum_{i = 1}^N \omega_i \sum_{j = 1}^N  \sin{\alpha[2\omega_j - 2\omega_i]} 
\end{multline}
where $\Sigma_\omega = \frac{1}{N}\sum_{j = 1}^N \omega_j^2$ denotes the estimate of the variance of the given set of natural frequencies. Considering the limit $N \rightarrow \infty$, with $\lim_{N \rightarrow \infty} \Sigma_\omega  = \sigma_{\omega}$ being the variance for the uniform frequency distribution with $d(\omega) = 0.5$ on the interval $[-1, 1]$, we arrive at
\begin{multline}\label{eq:oneClSol}
    \dot{\alpha} = 1 + \frac{\sin{\alpha}}{\alpha}\left( \frac{K_{1}}{\sigma_{\omega}}\left[\frac{\alpha\cos{\alpha}  - \sin{\alpha}}{\alpha^2} \right]\right)
    + \\
    \frac{\sin{\alpha}\cos{\alpha}}{\alpha}\left(\frac{K_{2}}{\sigma_{\omega}}\left[\frac{\alpha - \sin{\alpha}( \cos{\alpha} + 2\alpha\sin{\alpha})}{2\alpha^2} \right]\right).
\end{multline}
Note that the order parameter~\eqref{eq:defOrderParam} in the continuum limits becomes
\begin{align}\label{eq:orderParmCont1Cl}
    R_m = \left|\int_{-1}^1\frac{e^{\mathrm{i}m\alpha \omega}}{2}\,\mathrm{d}\omega\right|=\frac{\sin m\alpha}{m\alpha}.
\end{align}

To illustrate how the steady states for Eq.~\eqref{eq:oneClSol} behave, we plot the right hand side $g(\alpha)$ of Eq.~\eqref{eq:oneClSol} in for three different values of $K_1$ in Fig.~\ref{fig:1Cl_alphaSols}. In Fig.~\ref{fig:1Cl_alphaSols}, we observe that depending on the value of $K_1$ Eq.~\eqref{eq:oneClSol} give rise to two or no steady states corresponding to the roots of $g(\alpha)$. While continuously varying the coupling constant a pair of steady states may emerge in a saddle-node bifurcation. The stability of each steady state is determined by the derivative of $g(\alpha)$ at its roots and yields always one stable and one unstable state.

\begin{figure}
    \centering
    \includegraphics[width=\columnwidth]{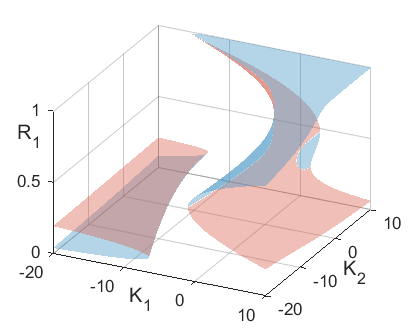}
    \caption{Order parameter $R_1$ (Eq.~\eqref{eq:orderParmCont1Cl}) corresponding to all steady state of Eq.~\eqref{eq:oneClSol} depending on the coupling constants $K_1$ and $K_2$. The blue and red shading of the surface correspond to stable and unstable states, respectively, with respect to Eq.~\eqref{eq:oneClSol}}
    \label{fig:1Cl_solsR1vsK1K2}
\end{figure}

In order to explore the solutions of Eq.~\eqref{eq:oneClSol} further, Fig.~\ref{fig:1Cl_solsR1vsK1K2} shows the values of the order parameter $R_1$ for each steady state depending on the coupling constants $K_1$ and $K_2$.

In section~\ref{sec:transitions}, we have shown transition scenarios for the cases $K_2=1.2$ and $K_2=2$. In Figure~\ref{fig:adiabatic}(a,c), the steady states for these two values of $K_2$ are displayed. We observe that for increasing value of $K_2$ the bifurcation lines extend towards the negative values of the coupling constant $K_1$ which agrees with the analytical findings.
\subsection{Approximation of phase-locked states of antipodal-type}\label{sec:collCooAntipodal}
Besides the emergence of in-phase-type phase-locked solution, we have also observed the appearance of another type, namely the antipodal-type, see Sec.~\ref{sec:transitions}. To describe approximately their existence in the framework of the collective coordinate approach, we consider two groups of oscillators $\phi^1_i$ and $\phi^2_i$ with sizes $N_1$ and $N_2$, respectively, where $N_1+N_2=N$. We assume that both groups evolve in coherence, however, with intergroup phase shift of $\pi$. Hence, for the collective coordinates we use
\begin{align*}
    \hat{\phi}^1_i &= \alpha(t) \omega^1_i, \\
    \hat{\phi}^2_i &= \alpha(t) \omega^2_i + \pi. 
\end{align*}
Using the same approach as in the section before, we derive the dynamical equations for the collective coordinate $\alpha(t)$. For brevity, we first introduce the following order parameters for each group $\mu= 1, 2$ as
\begin{align*}
    Z_m = n_{1} Z_m^{1} + (1-n_1)Z_m^{1}, 
\end{align*}
with $n_{\mu} = {N_{\mu}}/{N}$ and
\begin{align*}
    Z_m^{\mu} = \frac{1}{N_{\mu}}\sum_{j = 1}^{N_{\mu}} e^{\mathrm{i}m\phi_j^{\mu}}.
\end{align*}

We find that the error vector $\bm{\varepsilon}=(\bm{\varepsilon}^1,\bm{\varepsilon}^2)$ also splits up into two parts corresponding to the two groups. For the error we find
\begin{gather*}
    \varepsilon_k^{\mu} = \dot{\alpha}\omega_k^{\mu} - \omega_k^{\mu} - K_1 \text{Im}\left\{Z_1e^{-i\left(\alpha\omega_k^{\mu} + \psi^{\mu}\right)}\right\}\\ - K_2\text{Im}\left\{Z_2e^{-2\mathrm{i}\left(\alpha\omega_k^{\mu} + \psi^{\mu}\right)}\right\}
\end{gather*}
By again minimizing the error to reduce the synchronization problem~\eqref{eq:collCooErrMin}, we arrive at the following evolution equation in terms of the collective coordinate $\alpha$,
\begin{multline}\label{eq:cc_twoCl_finite}
    \dot{\alpha} = 1 + \frac{1}{\Sigma_\omega} \sum_{\mu = 1}^{2}\frac{n_{\mu}}{N_{\mu}} \sum_{j = 1}^{N_\mu}\omega_j^{\mu}\left[K_1 \text{Im}\left\{Z_1e^{-\mathrm{i}\left(\alpha\omega_j^{\mu} + \psi^{\mu}\right)}\right\}\right. \\
   \left. + K_2\text{Im}\left\{Z_2e^{-2\mathrm{i}\left(\alpha\omega_j^{\mu} + \psi^{\mu}\right)}\right\}\right]
\end{multline}
with $\psi^1=0$ and $\psi^2=\pi$, and $\Sigma_\omega=n_1\Sigma^1_\omega+n_2\Sigma^2_\omega$ being the estimates for the variance of the frequency distribution of the full set of oscillators and for the variance within the two individual clusters. Taking the limit $N_{\mu}\rightarrow\infty$ for all $\mu$, and hence $N\rightarrow\infty$, we have
\begin{multline}\label{eq:cc_twoCl_cont}
    \dot{\alpha} = 1 + \frac{1}{\sigma_\omega} \left[K_1 \text{Im}\left\{Z_1 I_1(\alpha)\right\} + K_2\text{Im}\left\{Z_2 I_2(\alpha)\right\}\right]
\end{multline}
where we define the resulting integrals as $I_m = n_{1}I^1_m + n_{2}I^2_m$ with $m=1,2$ and 
\begin{align}\label{eq:twoClusterIntDef}
   I_m^{\mu}(\alpha) &= \int d_\mu(\omega) \omega e^{-\mathrm{i}m\left(\alpha\omega + \psi^{\mu}\right)}\,\mathrm{d}\omega
\end{align}
with $d_\mu(\omega)$ denoting the distribution densities of the natural frequencies within the $\mu$th cluster, see App.~\ref{app:antipodal} for details on the explicit form of the continuum equations and their derivation.

\begin{figure}
    \centering
    \includegraphics[scale=0.34]{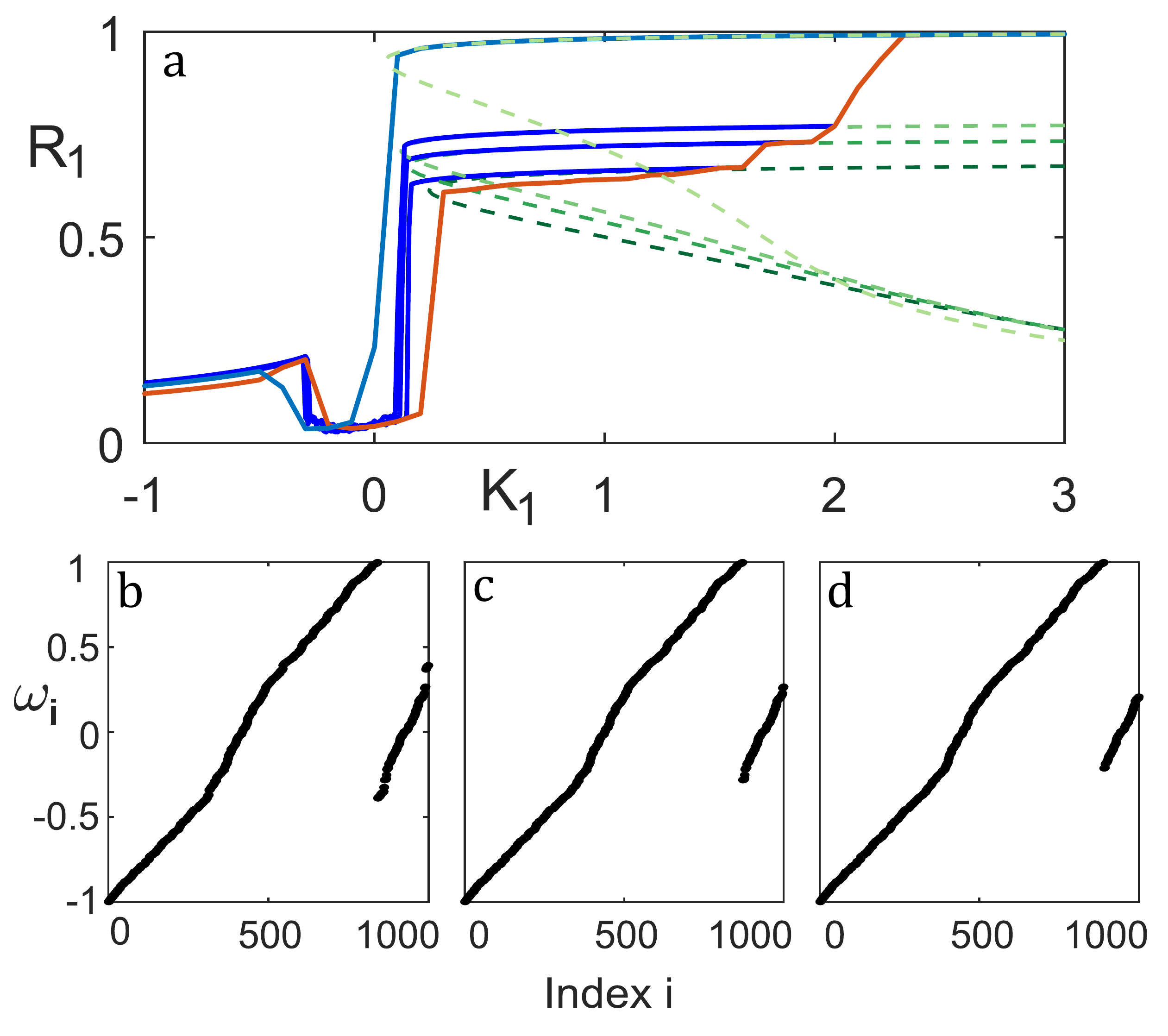}
    \caption{Top panel: Sweep-up (red) and sweep-down (blue) for $K_2 = 1.2$ with step size $\Delta K_1 = 0.1$ on the interval $[-20, 10]$, along with its analytical solution (lightest green line). Also shown are various sweep-downs (dark blue) that begin at certain target $K_1$ values compared against their analytical solutions for the target values of $K_1 = 1.5$,  $K_1 = 1.9$, and $K_1 = 2.0$ (darkest green, progressively lighter as target $K_1$ increases). Bottom panel: The corresponding natural frequency distributions for $K_1 = 1.5$,  $K_1 = 1.9$, and $K_1 = 2.0$, respectively (left to right).}
    \label{fig:2Cl_solsR1vsK1_syncTrans}
\end{figure}
In order to examine the derived approximations for the phase-locked states of antipodal-type, we plot solutions of Eq.~\ref{eq:cc_twoCl_cont} for three different configuration observed in the synchronization transition shown in Fig.~\ref{fig:snaps_k2_12}. In Figure~\ref{fig:2Cl_solsR1vsK1_syncTrans} the synchronization transition and the phase-locked states of antipodal-type for three different values of $K_1$ are displayed. In the following, we analyze their emergence. As described in Sec.~\ref{sec:transitions}, the number of oscillators in the first cluster $N_1$ increases with increasing $K_1$. For each of the selected values of $K_1$, we perform a sweep-up and sweep-down analysis starting each protocol with the antipodal-type states found in the synchronization transition. For the latter states, the natural frequencies are shown in Fig.~\ref{fig:2Cl_solsR1vsK1_syncTrans}(b-d) sorted by the phase of the oscillators. Taking the different values for $N_1$ of the three phase-locked states into account, we can show the steady state solutions of Eq.~\ref{eq:cc_twoCl_cont} alongside the results of the sweeping protocols. We observe that the states are well approximated by the collective coordinate approach for a wide range of coupling strengths $K_1$. Furthermore, their emergence for increasing values of $K_1$ is well described by the fold point of the solution curves to Eq.~\ref{eq:cc_twoCl_cont} (dashed lines). This is very similar to what we have already described for the phase-locked states of in-phase-type in Sec.~\ref{sec:collCoordInPhase}. 

The disappearance of the antipodal-type states for increasing $K_1$ is, however, not well captured by the collective coordinate ansatz. As we will see, the stability of the approximated solutions is not explaining the disappearance of the antipodal-type states. In fact, a closer numerical analysis (results not shown) reveals that the smaller cluster breaks up as $K_1$ increases. 
Moreover, close to the disappearance of the antipodal-type state the relation between the phase and the natural frequencies can not be described by a single collective coordinate (scaling factor) anymore but rather two scalings are necessary.
In order to accurately describe the disappearance, we generalize the collective coordinate ansatz as shown in the subsequent section.

\begin{figure}
    \centering
    \includegraphics[scale=0.34]{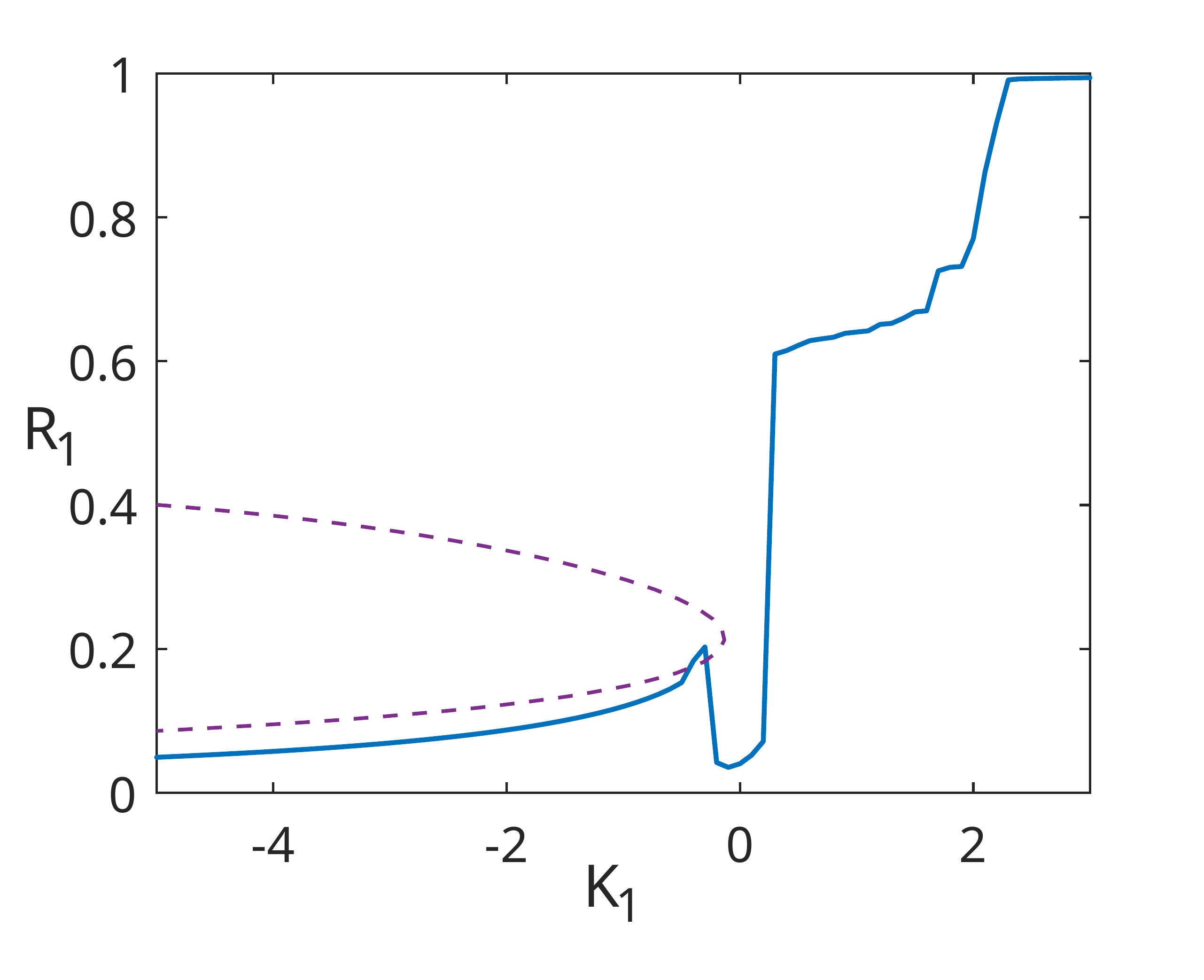}
    \caption{The negative branch analytical solution (purple) compared with the sweep-up (blue) for $K_2 = 1.2$ for continuous $K_1$ with step size $\Delta K_1 = 0.1$.}
    \label{fig:2Cl_solsR1vsK1_equalSize}
\end{figure}
Another phase-locked states of antipodal-type could be observed for negative values of the coupling strength $K_1$ in Fig.~\ref{fig:snaps_k2_12}(a), see also Fig.~\ref{fig:adiabatic}(a). Here, the distribution of the natural frequencies within each of the antipodal clusters is very different from the uniform distribution. Therefore, we use the form of the reduced equation for a finite frequency distribution~\eqref{eq:cc_twoCl_finite} in order to approximate the existence of this state. We further note that different to the antipodal-type states discussed before, the sizes of the clusters are almost equal and the average natural frequency of each cluster is different to zero. In Figure~\ref{fig:2Cl_solsR1vsK1_equalSize}, we show the result of the sweep-up from Fig.~\ref{fig:snaps_k2_12} together with the curve of solutions to Eq.~\ref{eq:cc_twoCl_finite}. Here, the collective coordinate ansatz approximates well the almost equally sized antipodal-type cluster. Also here this state disappears in a fold bifurcation. 
\subsection{Generalized two-cluster approach for phase-locked states of antipodal-type}\label{sec:collCooAntipodalGen}
In this section, we generalize the collective coordinate ansatz from the previous section in order to describe antipodal-type phase-locked states where two collective scaling factors $\alpha_1$ and $\alpha_2$ are necessary. We use the ansatz
\begin{align}\label{eq:cc_antipodal_gen}
    \hat{\phi}^1_i &= \alpha_1(t) \omega^1_i, \\
    \hat{\phi}^2_i &= \alpha_2(t) \omega^2_i + \pi. 
\end{align}
Following the general procedure described in Sec.~\ref{sec:collectiveCoord} and minimizing the error made by considering the reduced phase space, we arrive at
\begin{multline}\label{eq:cc_twoClGen_finite}
    \dot{\alpha}^\mu = 1 + \frac{1}{\Sigma^\mu_\omega} \frac{n_{\mu}}{N_{\mu}} \sum_{j = 1}^{N_\mu}\omega_j^{\mu}\left[K_1 \text{Im}\left\{Z_1e^{-\mathrm{i}\left(\alpha\omega_j^{\mu} + \psi^{\mu}\right)}\right\}\right. \\
   \left. + K_2\text{Im}\left\{Z_2e^{-2\mathrm{i}\left(\alpha\omega_j^{\mu} + \psi^{\mu}\right)}\right\}\right]
\end{multline}
where $\psi^1=0$, $\psi^2=\pi$ and $\Sigma_\omega^\mu$ is the estimate for the variance of the frequency distribution within the two individual clusters. The derivation of Eq.~\eqref{eq:cc_twoClGen_finite} follows analogously to the cases described in the previous two sections and we spare further details for the sake of brevity.

\begin{figure}
    \centering
    \includegraphics{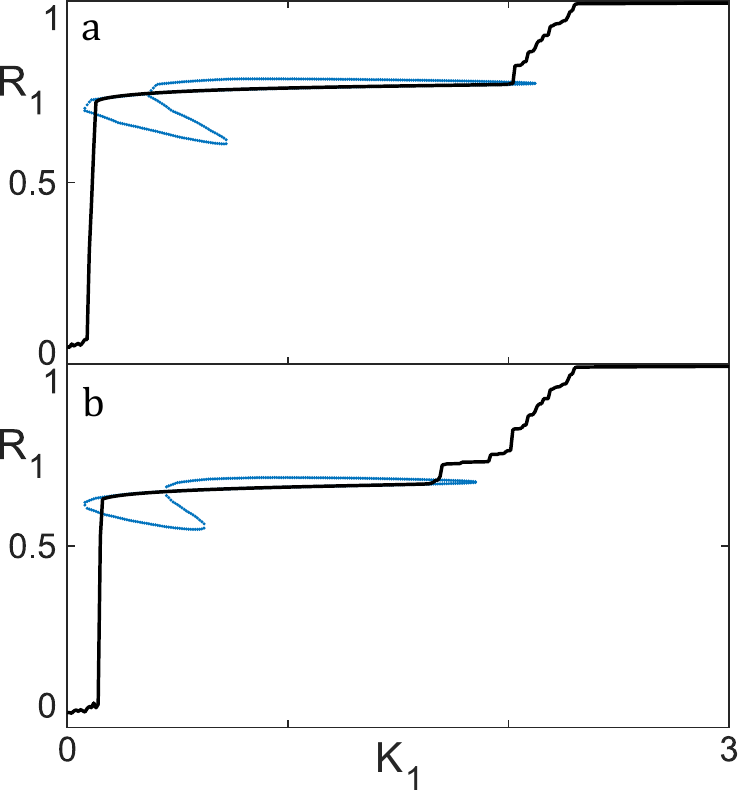}
    \caption{The sweep-up and sweep-down (black lines) for $K_2 = 1.2$ with step size $\Delta K_1 = 0.1$ along with the analytical solutions for Eq.~\eqref{eq:cc_twoClGen_finite} (blue lines). Pictured are the sweep-up and downs that begin at certain phase-locked antipodal-type states found in Fig.~\ref{fig:adiabatic}(a,b) for the values (a) $K_1 = 1.5$ and (b) $K_1 = 2$. The corresponding natural frequency distributions for $K_1 = 1.5$ and $K_1 = 2.0$ are displayed in Fig.~\ref{fig:2Cl_solsR1vsK1_syncTrans}(b,d), respectively.}
    \label{fig:2Cl_solsR1vsK1_syncTrans_gen}
\end{figure}
By using the reduced equations~\eqref{eq:cc_twoClGen_finite}, we can determine the existence of antipodal-type phase-locked states where each cluster possess a different phase scaling. In Fig.~\ref{fig:2Cl_solsR1vsK1_syncTrans_gen}, we show the numerical as well as analytical results for the existence of antipodal-type cluster states found in the transition displayed in Fig.~\ref{fig:snaps_k2_12}. It is seen that the generalized collective coordinate approach in Eq.~\eqref{eq:cc_antipodal_gen}, describes the emergence and disappearance of antipodal-type states depending on the values of $K_1$ very well. In fact, the antipodal-type cluster states emerge and disappear in saddle node bifurcations.

An exhaustive analysis of the solutions of Eqs.~\ref{eq:cc_twoClGen_finite} is beyond the scope of this work. We note that the complex solution curves displayed in Fig.~\ref{fig:2Cl_solsR1vsK1_syncTrans_gen} have to be considered as projections from two variables $(\alpha_1,\alpha_2)$ to one variable $R_1$. In fact, the crossing of the line in $R_1$ depending on the parameter $K_1$ with itself would not appear in the larger $(\alpha^1,\alpha^2)$ space. Therefore, the visible crossings do not indicate bifurcation points for the steady state solutions. Moreover, it is worth to mention that we have restricted our attention to states that are either of in-phase or antipodal-type. Two-cluster states where the clusters have a another phase difference than $\pi$ could be analyzed by e.g. considering an even more general collective coordinate ansatz such as $\hat{\phi}^1_i = \alpha_1(t) \omega^1_i,\hat{\phi}^2_i = \alpha_2(t) \omega^2_i + \psi(t)$ where $\psi(t)$ serve as an additional dynamical variable. Also, more than two clusters could be considered as they are shown in Fig.~\ref{fig:snaps_k2_2}(c,d) and Fig.~\ref{fig:adiabatic}(d).
\subsection{Approximations for the transverse stability of the one and two cluster states}\label{sec:stability}
In the sections~\ref{sec:collCoordInPhase}--\ref{sec:collCooAntipodalGen}, we have derived conditions for the existence of synchronous states of in-phase and antipodal-type. Such states had been discussed in Refs.~\onlinecite{KOM13a,KOM14} by using a self-consistency approach in the continuum limit, and existence results had been derived. However, as remarked in Refs.~\onlinecite{KOM13a,KOM14}, their approach is not capable of deriving stability conditions of the states transversally to the reduced subsystem. With the collective coordinate approach, we are in the fortunate position that this ansatz allows for an approximation of the stability of synchronous states. In particular, we consider a solution $\alpha$ of the collective coordinate description as stable if (i) $\alpha$ is stable with respect to the system Eq.~\eqref{eq:twoModeModel} reduced to the sub-manifold described by the collective coordinates, i.e., Eq.~\eqref{eq:dyn_coll_coord}, and (ii) the approximation $\hat{\phi}$ of the synchronized state is stable in the full model Eq.~\eqref{eq:twoModeModel}, see also~\onlinecite{SMI21}.

In order to derive some simple approximations for the stability of the synchronous states, we linearize the dynamical system~\eqref{eq:twoModeModel} around the approximated solution $\hat{\phi}(\alpha,\bm{\omega})$ which results in
\begin{align}
    \dot{\delta\bm{\phi}} = J(\alpha,\bm{\omega}) \delta\bm{\phi}
\end{align}
where $J$ denotes the Jacobian matrix with entries given by
\begin{align}\label{eq:JacobianEntries}
    J_{ij} &= \frac{K1}{N}\cos(\hat{\phi}_j - \hat{\phi}_i) + 2 \frac{K_2}{N}\cos2(\hat{\phi}_j-\hat{\phi}_i) \\
    J_{ii} &= - \left(\frac1N\sum_{j=1,j\ne i}^N K_1\cos(\hat{\phi}_j-\hat{\phi}_i) + 2 K_2\cos2(\hat{\phi}_j-\hat{\phi}_i)\right).
\end{align}

For the synchronous states of in-phase-type, we consider the (zeroth order) approximation $\hat{\phi}_j - \hat{\phi}_i\approx 0$ which results in $J_{ij} = \frac{1}{N}(K_1 + 2 {K_2})$ and $J_{ii} = - \frac{N-1}{N}(K_1 + 2 K_2)$ ($i\ne j$). For this Jacobian matrix the eigenvalues are given by $\lambda_0 = 0$ and $\lambda_1=-(K_1 + 2 K_2)$ (multiplicity of $N-1$). Hence, in this approximation the in-phase-type synchronous state is stable if it is stable with respect to the reduced equation~\eqref{eq:oneClSol} (or ~\eqref{eq:oneClSol_finite}) and if $K_1 + 2 K_2>0$.

For the synchronous states of antipodal-type, we have to divide the set of oscillators into two groups of size $N_1$ and $N_2$ labeled with $\mu=1,2$ and may consider the (zeroth order) approximation $\hat{\phi}^\mu_j - \hat{\phi}^\mu_i\approx 0$ and $\hat{\phi}^2_j - \hat{\phi}^1_i\approx 0$. Assuming this, the Jacobian also splits up into four blocks $J_{ij}^{\mu\nu}$, i.e.,
\begin{align*}
    J = \begin{pmatrix}
        J^{11} & j^{12}\hat{1}_{1,2}\\
        j^{21}\hat{1}_{2,1} & J^{22}
    \end{pmatrix}
\end{align*}
with $J^{\mu\mu}_{ii}=-\frac{N_\mu-1}{N}(K_1+2K_2)-\frac{N_\nu}{N}(2K_2-K_1)$, $J^{\mu\mu}_{ij}=\frac{1}{N}(K_1 + 2 {K_2})$, $j^{\mu\nu} = \frac{1}{N}(2 {K_2} - K_1)$  and $\hat{1}_{\mu\nu}$ being a matrix of size $N_\mu\times N_\nu$ ($\nu\ne\mu$, $i\ne j$) with all entries being $1$. According to Lemma A.2 in~\onlinecite{BER19a}, the eigenvalues of $J$ are given by $\lambda^1_1=-(2n_1-1)K_1-2K_2$ (multiplicity $N_1-1$), $\lambda^1_1=(2n_1-1)K_1-2K_2$ (multiplicity $N_2-1$), $\lambda_{0,1}=0$, $\lambda_{0,2}=K_1-2K_2$. Hence, in this approximation the antipodal-type synchronous state is stable if it is stable with respect to the reduced equation~\eqref{eq:cc_twoCl_cont} (or ~\eqref{eq:cc_twoCl_finite}) and if $\frac{-2K_2}{2n_1-1}<K_1<2K_2\le \frac{2K_2}{2n_1-1}$ since $n_1\ge 1/2$ without loss of generality.

In the following section, we wrap up our findings by looking at the transition scenarios for various values of $K_2$. We further show the implications of the stability analysis for the transitions observed. 
\section{Synchronization transitions in the $(K_1,K_2)$ plane}\label{sec:transitions_K1K2}
In section~\ref{sec:transitions}, we have discussed the transitions for an increasing coupling strength $K_1$ and two specific values of the coupling strength $K_2$. In this section, we show the transitions for a wide range of $K_2$.

\begin{figure}
    \centering
    \includegraphics{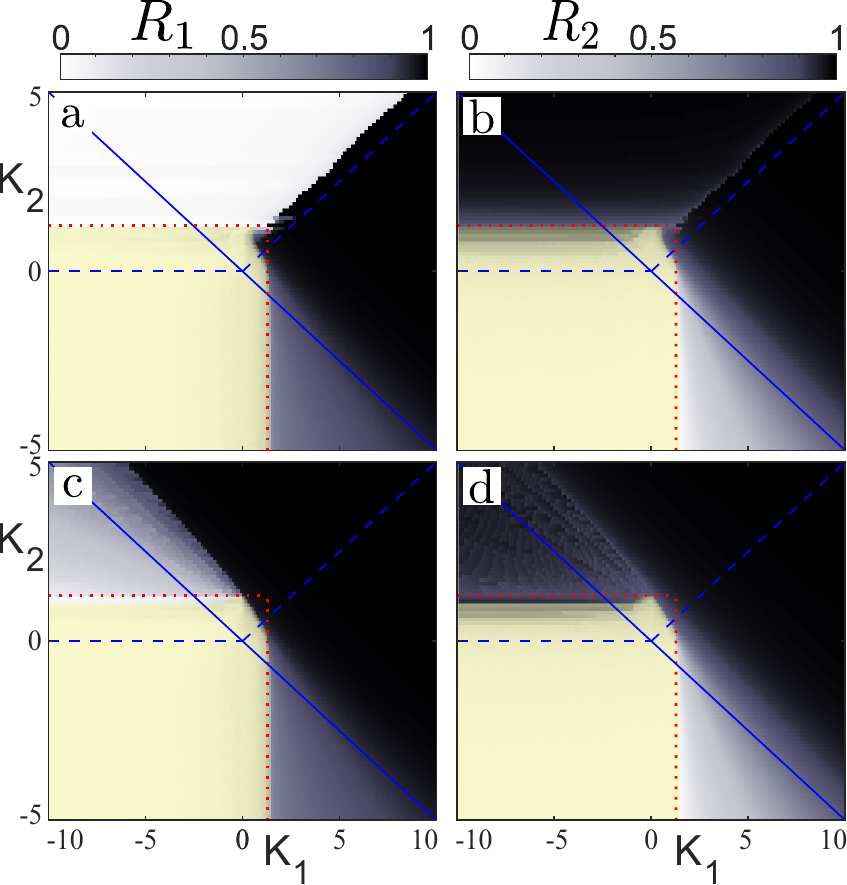}
    \caption{Up- and down-sweep for the first mode coupling strength $K_1$ for different values of the second mode coupling strength $K_2$ with step size $\Delta K_1 = 0.1$ and $\Delta K_2 = 0.1$. The gray scale shading represents the averaged values of the first (panels a,c) and second (panels b,d) order parameter $R_1$ and $R_2$, respectively. The yellow shaded areas indicate that the observed states are asynchronous, i.e. at least one average frequency $\Omega_i=1/T\int_{t_0}^{t_0+T} \dot{\phi}_i(t)\,\mathrm{d}t$ is different from all the others, where the threshold to measure the difference is set to $0.01$. For parameter values below the dotted red line, the incoherent states are stable, i.e., $K_{1,2}\le 2/\pi d(0)$. The blue line shows the boundary between regions where the synchronous state of in-phase-type is unstable (parameters in the lower left corner) or stable (parameters in the upper right corner). Similarly, the blue dashed line shows the boundary between regions where all synchronous states of antipodal-type are unstable (parameters in the upper left corner) or at least one state is stable (parameters in the lower right corner), see also Sec.~\ref{sec:stability} for the stability results. Procedure and all other parameters as in Fig.~\ref{fig:adiabatic}.}
    \label{fig:adiabaticSweep_R1R2vsK1vsK2}
\end{figure}

Figure~\ref{fig:adiabaticSweep_R1R2vsK1vsK2} shows the results of an adiabatic sweep-up and sweep-down continuation analysis for different values of $K_2$. We observe the presence of asynchronous dynamics for low values of $K_1$ and $K_2$ (yellow shaded region). The boundaries of this region are well described by the analytical stability result for incoherent states. Beyond the boundaries the dynamical system settles to different phase-locked states. Many of them are the described in the previous sections. In particular, for $K_2>0$ the transition scenarios are very similar to those described before in Sec.~\ref{sec:transitions}. Moreover, the multistability induced by the interplay of the two modes in the interaction function is also present in a wide range of parameters. Note that the multistability shown in the present analysis depends on the coupling strength $K_1$ which is the parameter that is changed during the adiabatic continuation. We further acknowledge that the abrupt (first-order) transition to full synchronization for an increasing coupling strength $K_1$, as described Ref.~\onlinecite{PAZ05a}, can be found for values $K_2>0$ up to almost the stability boundary $2/\pi d(0)\approx 1.27$.

Additionally, we note that the approximated stability boundaries for phase-locked states of in-phase and antipodal-type are qualitatively very good. More precisely, the conditions describe the maximal extension of the regions in which the corresponding states could be found. The differences to the numerically observed boundaries can be accounted, on the one hand, to the fact that the assumption $\hat{\phi}_j^\mu-\hat{\phi}_j^\mu\approx 0$ becomes bad close the saddle node bifurcation (drop in in the value of $R_1$, see Fig.~\ref{fig:adiabatic}). On the other hand, the stability condition does not imply existence and hence the line of saddle-node bifurcations not necessarily coincides with the stability boundary, see Secs.~\ref{sec:collCoordInPhase}--\ref{sec:collCooAntipodalGen} for the existence results.

For coupling strengths $K_2<0$, the transition scenarios change. In fact, we observe a sudden transition between asynchrony to synchrony close to the stability boundary for the incoherent state. As in the case $K_2>2/\pi d(0)$, there are no direct transitions to full synchrony but again to phase-locked states that eventually turn into synchronous states of in-phase-type for very large $K_1$. Remarkably, these transition scenarios show no hysteresis, compare e.g. the up- and down-sweep protocols shown in Fig.~\ref{fig:adiabaticSweep_R1R2vsK1vsK2}a,c. Moreover, the phase-locked states observed in the transition still show a separation into approximately two groups with a phase shift of $\pi$ in their collective phases of the first moment. However, the structure within the groups can be very different and even further split up into more groups. Due to this additional complexity in the transition, we skip any further detailed analysis.

%
%
\section{Conclusions}\label{sec:conclusion}
In this work, we have studied the synchronization transition in a phase oscillator system with a two-mode interaction function. We have shown that the transition scenarios crucially depend on the interplay of the two considered modes. Using adiabatic continuation protocols, we have described parameter regimes with a high degree of multistability of phase-locked states of in-phase and antipodal-type. The multistable states, in addition to giving rise to hysteresis, allow for a cascade of first-order transitions to synchrony, similar single-step transition recently found for adaptive dynamical networks~\cite{FIA22}.

In order to describe the existence of the multiple phase-locked states, we have extended the collective coordinate approach~\cite{GOT15} to system with multi-mode interaction. Using our extended approach, we are able to characterize the emergence of the observed states through saddle-node bifurcations and approximate their regions of existence very well.

In essence of our theoretical analysis, we have shown that the interplay of two modes may lead to the emergence of two independent groups of oscillators whose interaction determine the final dynamical states. This property of the systems has been particularly important to describe the disappearance of certain phase-locked states of antipodal-type where two independent collective coordinates had to be used for a proper analysis. In comparison to single-mode interaction functions, the emergence of two independently interacting groups represents the higher degree of dynamical complexity due to two modes in the interaction function.

Phase oscillator models are a commonly used paradigm to study synchronization phenomena in systems of coupled oscillators~\cite{ACE05}. In particular, phase reduction techniques have shown that interacting dynamcial systems which are close to a Hopf bifurcation can be well described by phase oscillator models~\cite{PIE19a}. However, the interaction function of these phase oscillator models might have a complicated form and is generally different from the standard Kuramoto-type single-mode coupling. Additionally, in works on the phase reduction beyond the weak coupling limit, mixed-mode couplings appear quite naturally in the phase model approximations~\cite{ASH16b,LEO19,GEN20}. In this work, we have shown, in which way the interplay of different modes in the interaction function lead to more complex dynamical scenarios. Hence, our analysis provides a perspective for future research on dynamical systems including complex interaction functions beyond the Kuramoto-type.


\section*{Data Availability Statement}
The data that supports the findings of this study are available within the article.
\appendix
\section{Derivation of collective coordinate approximation for states of in-phase-type}\label{app:inPhase}
In this section, we derive the equation of motion for the collective coordinate approximation for states of in-phase-type. As given in Eq.~\eqref{eq:oneCluster_error}, the error function reads
\begin{multline*}
     \varepsilon_i 
     = \dot{\alpha} \frac{d\hat{\phi}_i}{d\alpha} - \omega_i - \frac{K_{1,1}}{N}\sum_{j = 1}^N \sin{\alpha[\omega_j - \omega_i]}\\
     -\frac{K_{1,2}}{N}\sum_{j = 1}^N \sin{2\alpha[\omega_j - \omega_i]}
\end{multline*}
Now using the condition that $\langle \bm{\varepsilon}, \frac{d\hat{\bm{\phi}}}{d\alpha}\rangle = 0$, and noting that $\frac{d\hat{\phi}_i}{d\alpha} = \omega_i$, we obtain
\begin{multline*}
    0=\langle \bm{\varepsilon}, \frac{d\theta}{d\alpha}\rangle = \dot{\alpha} \langle \frac{d\hat{\bm{\phi}}}{d\alpha}, \frac{d\hat{\bm{\phi}}}{d\alpha}\rangle - \langle \frac{d\hat{\bm{\phi}}}{d\alpha}, \frac{d\hat{\bm{\phi}}}{d\alpha}\rangle \\
    - \frac{K_1}{N}\sum_{i,j = 1}^N \sin{\alpha[\omega_j - \omega_i]} \frac{d\hat{\phi}_i}{d\alpha}\\
     -\frac{K_2}{N}\sum_{i,j = 1}^N \sin{2\alpha[\omega_j - \omega_i]} \frac{d\hat{\phi}_i}{d\alpha}
\end{multline*}
This equation yields 
\begin{multline}
    \dot{\alpha} = 1 + \frac{K_{1}}{\Sigma_\omega} \frac{1}{N^2}\sum_{i = 1}^N \omega_i \sum_{j = 1}^N  \sin{\alpha[\omega_j - \omega_i]}  \\
    +\frac{K_{2}}{\Sigma_\omega} \frac{1}{N^2}\sum_{i = 1}^N \omega_i \sum_{j = 1}^N  \sin{\alpha[2\omega_j - 2\omega_i]} 
\end{multline}
where $\Sigma_\omega = \frac{1}{N}\sum_{j = 1}^N \omega_j^2$.
Considering a uniform distribution of natural frequencies on the interval $[-1,1]$ with the distribution density $d(\omega) = 0.5$ and taking the continuum limit as $N \rightarrow \infty$ we obtain
\begin{gather*}
    \dot{\alpha} = 1 + \frac{K_{1,1}}{\sigma_{\omega}^2}  \int_{-1}^1 \frac{\omega}{4} \int_{-1}^1 \sin{\alpha[\eta - \omega]} \,d\eta \,d\omega\\
    + \frac{K_{1,2}}{\sigma_{\omega}^2}  \int_{-1}^1 \frac{\omega}{4} \int_{-1}^1 \sin{2\alpha[\eta - \omega]} \,d\eta ,d\omega
\end{gather*}
where we denote $\lim_{N\to\infty}\Sigma_\omega = \sigma_\omega$.

\begin{figure}
    \centering
    \includegraphics{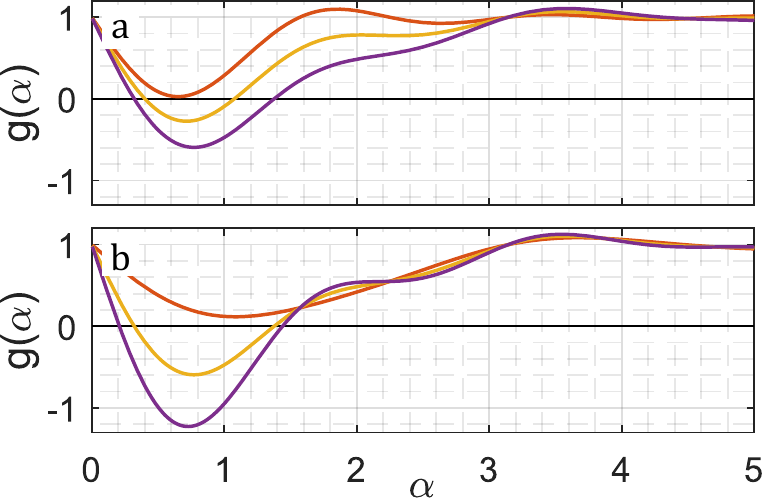}
    \caption{Plots for the function $g(\alpha)$ on the right-hand side of Eq.~\eqref{eq:oneClSol}. Panel (a) displays the function depending on $K_1$ for $K_1=0$ (red), $K_1=0.5$ (yellow), $K_1=1$ (blue) and fixed $K_2=1.2$. Panel (b) displays the function depending on $K_2$ for $K_2=0.2$ (red), $K_2=1.2$ (yellow), $K_2=2$ (blue) and fixed $K_1=1$.}
    \label{fig:1Cl_alphaSols}
\end{figure}

After evaluating the integrals, we finally obtain
\begin{gather*}
    \dot{\alpha} = 1 + \frac{\sin{\alpha}}{\alpha}\left( \frac{K_{1}}{\sigma_{\omega}}\left[\frac{\alpha\cos{\alpha}  - \sin{\alpha}}{\alpha^2} \right]\right)
    \\ 
    +\frac{\sin{\alpha}\cos{\alpha}}{\alpha}\left(\frac{K_{2}}{\sigma_{\omega}}\left[\frac{\alpha - \sin{\alpha}( \cos{\alpha} + 2\alpha\sin{\alpha})}{2\alpha^2} \right]\right),
\end{gather*}
where $\sigma_\omega=1/3$ for the given frequency distribution, see also Eq.~\eqref{eq:oneClSol}. Figure~\ref{fig:1Cl_alphaSols} display the function $g(\alpha)$ on the right hand side of Eq.~\eqref{eq:oneClSol} depending on the coupling strengths $K_1$ and $K_2$. Note that crossings of the colored lines with the black line ($g(\alpha)=0$) correspond to steady states of the reduced equations. 

\section{Derivation of collective coordinate approximation for states of antipodal-type}\label{app:antipodal}
For states of antipodal-type, let us derive the two cluster solution. Recall the error which reads 
\begin{gather*}
    \varepsilon_i^{\mu} = \dot{\alpha}\omega_i^{\mu} - \omega_i^{\mu} - K_1 \text{Im}\left\{Z_1 e^{-\mathrm{i}\left(\alpha^{\mu}\omega_i^{\mu} + \psi^{\mu}\right)}\right\}\\ - K_2\text{Im}\left\{Z_2 e^{-2\mathrm{i}\left(\alpha^{\mu}\omega_i^{\mu} + \psi^{\mu}\right)}\right\}
\end{gather*}
where we consider the general collective coordinate ansatz $\hat{\phi}_i^\mu=\alpha\omega^\mu_i + \psi^\mu$ for $\psi^\mu\in[0,2\pi)$, $\mu=1,2$ and $k=1,\dots,N^\mu$. Taking $\langle \bm{\varepsilon}, \frac{\mathrm{d}\bm{\hat{\phi}}}{\mathrm{d}\alpha}\rangle = 0$, we obtain
\begin{multline*}
    \dot{\alpha} = 1 + \frac{1}{\Sigma_\omega}\sum_{\mu = 1}^{2}\frac{n_{\mu}}{N_{\mu}} \sum_{j = 1}^{N_\mu}\omega_j^{\mu}\left[K_1 \text{Im}\left\{Z_1e^{-\mathrm{i}\left(\alpha\omega_j^{\mu} + \psi^{\mu}\right)}\right\}\right. \\
   \left. + K_2\text{Im}\left\{Z_2e^{-2\mathrm{i}\left(\alpha\omega_j^{\mu} + \psi^{\mu}\right)}\right\}\right]
\end{multline*}
with
\begin{align*}
    \Sigma_\omega = \frac{1}{N}\sum_{\mu = 1}^{2} \sum_{j = 1}^{N_\mu}\left(\omega_j^{\mu}\right)^2.
\end{align*}
Taking the limit $N_{\mu}\to\infty$ for all $\mu$ and consequently $N\to\infty$, we get
\begin{multline*}
   \frac{1}{N_{\mu}}\sum_{j = 1}^{N_\mu}\omega_j^{\mu}K_1 \text{Im}\left\{Z_1e^{-\mathrm{i}\left(\alpha\omega_j^{\mu} + \psi^{\mu}\right)}\right\} \\
    \to K_1 \text{Im}\left\{Z_1\int d^\mu(\omega) \omega e^{-\mathrm{i}\left(\alpha\omega + \psi^{\mu}\right)}\,\mathrm{d}\omega\right\} 
\end{multline*}
and similarly for the $K_2$ term with $d^\mu(\omega)$ being the distribution densities of natural frequencies of the $\mu$th cluster. Thus it can be seen that the equation becomes
\begin{multline*}
     \dot{\alpha} = 1 + \frac{1}{\sigma_\omega}\sum_{\mu = 1}^{\eta}n_{\mu}\left[K_1 \text{Im}\left\{Z_1 I_1^{\mu}\right\} + K_2\text{Im}\left\{Z_2I_2^{\mu}\right\}\right]
\end{multline*}
where $I_m^\mu$ are defined as in Eqs.~\eqref{eq:twoClusterIntDef} and $\Sigma_\omega \to \sigma_\omega$ as $N\to\infty$.

In agreement with the observations described in Sec.~\ref{sec:transitions}, we assume that the natural frequencies within each cluster $\mu=1,2$ are uniformly distributed on an interval $[a_1^{\mu}, a_2^{\mu}]$. For the distribution densities, we assume $d_\mu(\omega)=\frac{1}{2\Delta^\mu}$ on the interval $[a^\mu_1,a^\mu_2]$ for $a^\mu_1,a^\mu_2\in\mathbb{R}$ and $d_\mu(\omega)=0$ everywhere else, where we define $\Delta^\mu=\frac{a^\mu_2-a^\mu_1}{2}$ and $\bar{\omega}^\mu=\frac{a^\mu_1+a^\mu_2}{2}$. Then for two clusters with $\psi^1 = 0$ and $\psi^2 = \pi$ the order parameter can be written as
\begin{align*}
    Z_m &= n_1 Z_m^1 + (1-n_1) Z_m^2.
\end{align*}
For the individual order parameters of the cluster $Z_m^{\mu}$, we can derive the following integrals:
\begin{align*}
    Z_m^{\mu}
    &= \int_{a_1^\mu}^{a_2^\mu} d^\mu(\omega) e^{\mathrm{i}m(\alpha\omega+\psi^\mu)}\,\mathrm{d}\omega\\
    &= \frac{e^{\mathrm{i}m(\alpha\bar{\omega}^\mu+\psi^\mu)}}{2\Delta^\mu} \int_{-\Delta^\mu}^{\Delta^\mu} e^{\mathrm{i}m\alpha\omega}\,\mathrm{d}\omega \\
    &= \frac{e^{\mathrm{i}m(\alpha\bar{\omega}^\mu+\psi^\mu)}}{m\alpha\Delta^\mu}\sin(m\alpha\Delta^\mu)
\end{align*}
Here, we use the transformation $\omega\to \omega - \bar{\omega}^\mu$ in the second line. We finally obtain
\begin{align*}
    Z_m^\mu &=  \frac{(-1)^{m(\mu-1)}e^{\mathrm{i}m\alpha\bar{\omega}^\mu}}{m \alpha \Delta^\mu}\sin{(m \alpha \Delta^\mu)}.
\end{align*}
For the integrals $I^\mu_m = n_1 I_m^1 + (1-n_1) I_m^2$ we proceed analogously. Using the transformation $\omega\to \omega - \bar{\omega}^\mu$, we find
\begin{align*}
    I_\mu^{m} = \frac{(-1)^{m(\mu-1)}e^{-\mathrm{i}m\alpha\bar{\omega}^\mu}}{2\Delta^\mu}\int_{-\Delta^\mu}^{\Delta^\mu} \omega e^{-\mathrm{i}(\alpha\omega^{\mu})}\,\mathrm{d}\omega + \bar{\omega}^\mu (Z_m^\mu)^*.
\end{align*}
where the asterisk denotes the complex conjugate. Applying integration by parts, we finally obtain
\begin{align*}
    I_m^\mu = \hat{I}_m^\mu + \bar{\omega}^{\mu}({Z_m^{\mu})^*}
\end{align*}
with
\begin{align*}
    \hat{I}_m^\mu = (-1)^{m(\mu-1)}\mathrm{i}e^{-\mathrm{i}m\alpha\bar{\omega}^\mu}\frac{m\alpha\Delta^{\mu}\cos(m\alpha\Delta^{\mu}) - \sin(m\alpha\Delta^{\mu})}{m^2\alpha^2\Delta^{\mu}}.
\end{align*}

For the variance, we obtain
\begin{align*}
    \sigma_\omega = \frac{n_1((\Delta^1)^2+2(\bar{\omega}^1)^2)+(1-n_1)((\Delta^2)^2+2(\bar{\omega}^2)^2)}{3}=\frac13.
\end{align*}
\section*{References}

\end{document}